\documentstyle[12pt,epsf]{article}
\catcode`\@=11
\def\eqnarray{\let\@currentlabel=\theequation\refstepcounter{equation}
    \global\@eqnswtrue
    \global\@eqcnt\z@\tabskip\@centering\let\\=\@eqncr
    $$\halign to \displaywidth\bgroup\@eqnsel\hskip\@centering
      $\displaystyle\tabskip\z@{##}$&\global\@eqcnt\@ne 
       \hfil${{}##{}}$\hfil
      &\global\@eqcnt\tw@ $\displaystyle\tabskip\z@{##}$\hfil 
       \tabskip\@centering&\llap{##}\tabskip\z@\cr}
\def\lefteqn#1{\hbox to 4\arraycolsep{$\displaystyle #1$\hss}}

\newcommand{\prepr}[1] {\begin{flushright}  {\bf #1} \end{flushright} \vskip
1.cm}
\newcommand{\titul}[1] {\begin{center}{\Large {\bf #1 } } \end{center}
\vskip 0.8cm}

\newcommand{\autor}[1] {\begin{center}  {\bf \lineskip .3cm #1 }
                         \end{center} }

\newcommand{\where}[1] {\begin{center}  {\normalsize \bf \it #1   } \end{center}}

\newcommand{\abstr}[1] {{\begin{center} \vskip .5cm {\bf \large Abstract
                        \vspace{0pt}} \end{center}}\begin{quote} \small #1
                        \end{quote}}
\newcounter{muni}

\begin{document}
\begin{titlepage}
\prepr{Preprint/APCTP-98-03}

\titul{Factorization in Color-Suppressed B Meson Decays
\footnote{Contributed paper to BaBar Yellow Book} }
\autor{Y. Y. Keum}
\where{\rm Asia Pacific Center for Theoretical Physics \\
 207-43 Cheongryangri-Dong Dongdaemun-Gu, Seoul 130-012, Korea }

\thispagestyle{empty}
\abstr{
We summerize the status of factorization hypothesis in the
color-suppressed B meson decays : $B \rightarrow J/\psi K^{(*)}$.
We present the general formalism for decay rates and polarization
fractions with considering all possible non-factorizable contributions
which also include the color-octet contributions, and a new factorization
scheme comes appear when the universality is exist :
$\chi_{F1} = \chi_{A1} = \chi_{A2} = \chi_{V} = \chi$.
We consider various phenomenological models to compare their theoretical
predictions with the recent CLEO II experimental measurements.}
\end{titlepage}
\newpage
\section{Introduction}

One of the interests in $B \rightarrow J/\psi K^{*}$ decays is their
role in CP violation measurements at asymmetric B-factories.
The vector-vector decay $B^{0} \rightarrow J/\psi K^{*0}$, with
$K^{*0} \rightarrow K^{0}_s \pi^{0}$, is a mixture of CP-even and
CP-odd eigenstates since it can proceed via an S, P, D wave decay.
If one CP eigenstate dominates or if the two CP eigenstates can be 
separated, this decay can be used to measure the angle $\beta$ of 
the unitarity triangle in a manner similar to which the CP-odd 
eigenstates $B^{0} \rightarrow J/\psi K^{0}_S$ is used.

Measurements of the decay amplitudes of $B \rightarrow J/\psi K^{(*)}$
transitions also provide a test of the factorization hypothesis in decays 
with internal W-emission, so called Color-Suppressed decay modes.
Several phenomeological models, based on the factorization hypothesis,
predict the logitudinal polarization fraction in 
$B \rightarrow J/\psi K^{*}$, denoted $\Gamma_L/\Gamma$, and 
the ratio of vector to psedoscalar meson production,
$R_{\psi} \equiv {\cal{B}}(B\rightarrow J/\psi K^{*})/{\cal{B}}
(B \rightarrow J/\psi K)$
\cite{WBS,NR,ISGW,CDDFGN,Orsay}.
It has been noted \cite{Orsay,Gourdin} that usual form factor models
can not simultaneously explain the earilier experimental data for these
two quantities. As shown in table 1,
the high values of $\Gamma_L/\Gamma$ measured by ARGUS \cite{ARGUS}
and CLEO II \cite{CLEO2}, with low statistics, are not consistent with
factorization and the measured value of $R$.
The CDF collaboration has measured a lower value of $\Gamma_L/\Gamma$
\cite{CDF1}. Additional information about the validity of factorization
can be obtained by a measurement of the decay amplitude phases, since
any non-trivial phase differences indicate final state interactions
and the breakdown of factorization \cite{kroner}.
In recent CLEO collaboration \cite{CLEO3} presented a complete angular
analysis and an update of the branching fractions for $B \rightarrow
J/\psi K^{*0}$ using the full CLEO II data sample. They measured 
five quantities include $\Gamma_L/\Gamma = 0.52 \pm 0.07 \pm 0.04$,
and $R_{\psi} = 1.45 \pm 0.20 \pm 0.17$.
From the data of the relative phases $\phi(A_{\perp}), \phi(A_{\parallel})$
with respect to $\phi(A_{0})$, the amplitudes are relatively real,
and there is no significant signature of the final state interaction.

\section{General formalism for Decay rates and Polarization
in Color-suppressed Decay Modes}

Using the effective Hamiltonian that contains the short distance wilson
coefficients $C_1$ and $C_2$, the decay amplitude for such processes in 
written as
\begin{eqnarray}
& & {\cal A}(B \rightarrow P(V)J/\psi) 
= \langle P(V)J/\psi|{\cal H}^{eff}|B \rangle  \nonumber \\
& & \hspace{20mm} = {G_F \over \sqrt{2}} V_{cb}^{*} V_{cs} \,\,
\Bigl [ a_2(\mu)  \langle P(V)J/\psi|O^{(1)}|B \rangle 
+ 2 C_1(\mu) \langle P(V)J/\psi|O^{(8)}|B \rangle \Bigr ]
\label{chap10:cs01}
\end{eqnarray}
where
\begin{eqnarray}\label{chap10:cs02}
O^{(1)} &=& \bar c_i \gamma_\mu (1 - \gamma_5) c^i ~
\bar s_j \gamma^\mu (1 - \gamma_5) b^j, \nonumber \\
O^{(8)} &=& \bar c_i \gamma_\mu (1 - \gamma_5)
\frac{\lambda^l_{ij}}{2} c^j ~
\bar s_k \gamma^\mu (1 - \gamma_5)
\frac{\lambda^l_{kl}}{2} b^l
\end{eqnarray}
and
\begin{equation}\label{chap10:cs03}
a_2 = {C_1 \over N_c} + C_2
\end{equation}
$N_c$ is the number of colors and $\lambda^a$ is the Gell-Mann matrices.

Here we consider all possible non-factorizable contributions 
in Eq.(\ref{chap10:cs01}) and parameterize them as the following:
\begin{eqnarray}
\langle P(V)J/\psi|O^{(1)}|B \rangle &=&
\langle J/\psi|(\bar{c}c)_{V-A}|0 \rangle
\langle P(V)|(\bar{b}s)_{V-A}|B \rangle \nonumber \\
& & \hspace{10mm} + \hspace{3mm} 
\langle P(V)J/\psi|O^{(1)}|B \rangle_{NF}, 
\label{chap10:cs04} \\
 \langle P(V)J/\psi|O^{(8)}|B \rangle &=& 
 \langle P(V)J/\psi|O^{(8)}|B \rangle_{NF}.
\label{chap10:cs05}
\end{eqnarray}
and
\begin{eqnarray}
\langle J/\psi|(\bar{c}c)_{V-A}|0 \rangle &=& 
{\epsilon}^{\mu} m_{\psi} f_{\psi}
\label{chap10:cs06} \\
 \langle P |(\bar{b}s)_{V-A}|B \rangle &=&
(p_B^{\mu} + p_P^{\mu} - {m_B^2 - m_P^2 \over q^2}q^{\mu}) F_1(q^2) 
+ {(m_B^2 - m_P^2) \over q^2} q^{\mu} F_0(q^2),
\label{chap10:cs07} \\
\langle V|(\bar{b}s)_{V-A}|B \rangle &=&
- \Big[ (m_B + m_V) \eta^{*}_{\mu} A_1(q^2) -
{\eta^{*} \cdot q \over (m_B + m_V)}(p_B + p_V)_{\mu}A_2(q^2) \nonumber \\
& &  -2 m_V {\eta^{*} \cdot q \over q^2} q_{\mu} (A_3(q^2) - A_0(q^2)) 
 - {2 i \over m_B + m_V} \epsilon_{\mu \nu \rho \sigma}
\eta^{*\nu} p_B^{\rho} p_V^{\sigma} V(q^2) \Bigr ]
\label{chap10:cs08} 
\end{eqnarray}
\begin{equation}
\langle PJ/\psi|O^{(1,8)}|B \rangle_{NF} = 2 m_{\psi}f_{\psi}(\epsilon \cdot p_B)
F_1^{(1,8)NF}(q^2) 
\label{chap10:cs09} 
\end{equation}
\begin{eqnarray}
\langle VJ/\psi|O^{(1,8)}|B \rangle_{NF} &=& - m_{\psi} f_{\psi}
\Big[ (m_B + m_V) (\epsilon \cdot \eta^{*}) A_1^{(1,8)NF}(q^2)  
\nonumber \\
& &  -{2 \over m_B + m_V} (\epsilon \cdot p_B)(\eta^{*}\cdot p_B)
 A_2^{(1,8)NF}(q^2) \nonumber \\
& &  - {2i \over m_B + m_V} \epsilon_{\mu\nu\rho\sigma}
\epsilon^{\mu}\eta^{*\nu}p_B^{\rho}p_V^{\sigma} V^{(1,8)NF}(q^2) \Bigr ]
\label{chap10:cs10}
\end{eqnarray}
where $q_{\mu} = \left[p_B - p_{P(V)} \right]_{\mu} = (p_{\psi})_{\mu}$.
The polarization vectors $\epsilon^{\mu}$ and $\eta^{\mu}$ correspond to
the two vector mesons $J/\psi$ and $V$, respectively.

Substituting (\ref{chap10:cs04} -\ref{chap10:cs10}) into the decay amplitude
(\ref{chap10:cs01}), we can calculate decay rates for the processes 
$B \rightarrow P(V)J/\psi$ and polarization for the $B \rightarrow VJ/\psi$
process.

The decay widths for each process are presented below :
\begin{eqnarray}
\Gamma(B \rightarrow PJ/\psi) &=&
{G_F^2 m_B^5 \over 32 \pi} |V_{cb}|^2 |V_{cs}|^2 a_2^2
\left( {f_{\psi} \over m_B } \right)^2
k^3(t^2) \left|F_1(m_{\psi}^2)\right|^2 
\left|1 + {2C_1 \over a_2}\chi_{F1}\right|^2,
\label{chap10:cs11} \\
\Gamma(B \rightarrow VJ/\psi) &=&
{G_F^2 m_B^5 \over 32 \pi} |V_{cb}|^2 |V_{cs}|^2 a_2^2
\left( {f_{\psi} \over m_B } \right)^2
\left|A_1(m_{\psi}^2)\right|^2 
k(t^2)t^2(1+r)^2 \sum_{\lambda\lambda}H_{\lambda\lambda},
\label{chap10:cs12}
\end{eqnarray}
where
\begin{eqnarray}
H_{L} = H_{00} &=&
\Bigl [ a \left(1 + 2{C_1 \over a_2}\chi_{A1} \right) 
- b x \left(1 + 2{C_1 \over a_2}\chi_{A2} \right) \Bigr ]^2,
\label{chap10:cs13} \\
H_{T} = H_{++} + H_{--} &=&
2 \Bigl [ \left(1 + 2{C_1 \over a_2}\chi_{A1} \right)^2
+ c^2 y^2 \left(1 + 2{C_1 \over a_2}\chi_{V} \right)^2 \Bigr ],
\label{chap10:cs14} \\
\chi_{F1} &=& \left(F_1^{(8)NF}(m_{\psi^2}) 
+ {a_2 \over 2 C_1}F_1^{(1)NF}(m_{\psi^2})\right)/F_1(m_{\psi^2}),
\label{chap10:cs15} \\
\chi_{A1} &=& \left(A_1^{(8)NF}(m_{\psi^2}) 
+ {a_2 \over 2 C_1}A_1^{(1)NF}(m_{\psi^2})\right)/A_1(m_{\psi^2}),
\label{chap10:cs16} \\
\chi_{A2} &=& \left(A_2^{(8)NF}(m_{\psi^2}) 
+ {a_2 \over 2 C_1}A_2^{(1)NF}(m_{\psi^2})\right)/A_2(m_{\psi^2}),
\label{chap10:cs17} \\
\chi_{V} &=& \left(V^{(8)NF}(m_{\psi^2}) 
+ {a_2 \over 2 C_1}V^{(1)NF}(m_{\psi^2})\right)/V(m_{\psi^2}),
\label{chap10:cs18} 
\end{eqnarray}
Here subscripts $L$ and $T$ in (\ref{chap10:cs13})
and (\ref{chap10:cs14}) stand for {\it longitudinal}
and {\it transverse}, and $00, ++$ and $--$ represent the vector meson
helicities. 
In (\ref{chap10:cs11})- (\ref{chap10:cs14}) we have intorduce
the following dimensionless parameters:
\begin{equation}\label{chap10:cs19}
r = {m_{P(V)} \over m_B}, \hspace{15mm}
t = {m_{\psi} \over m_B},
\end{equation}
\begin{equation}\label{chap10:cs20}
k(t^2) = \sqrt{(1 - r^2 -t^2)^2 - 4 r^2 t^2},
\end{equation}
\begin{equation}
\label{chap10:cs21}
a = {1 -r^2 -t^2 \over 2 r t}, \hspace{10mm}
b = {k^2(t^2) \over 2 r t (1 + r)^2}, \hspace{10mm}
c = {k(t^2) \over (1 + r)^2}.
\end{equation}
The numerical values of parameters a, b, and c for the processes
$B \rightarrow J/\psi K(K^{*})$ are given as
\begin{equation}\label{chap10:cs22}
a = 3.165, \hspace{10mm}
b = 1.308, \hspace{10mm}
c = 0.436.
\end{equation}
Futhermore x, y, and z represent the following ratios,
\begin{equation}\label{chap10:cs23}
x = {A_2^{BK^{*}}(m_\psi^2) \over A_1^{BK^{*}}(m_\psi^2)}, \hspace{10mm}
y = {V^{BK^{*}}(m_\psi^2) \over A_1^{BK^{*}}(m_\psi^2)}, \hspace{10mm}
z = {F_1^{BK}(m_\psi^2) \over A_1^{BK^{*}}(m_\psi^2)}.
\end{equation}
The longitudinal polarization fraction $\Gamma_L/\Gamma$
 and the ratio $R_{\psi}$ are defined:
\begin{equation}\label{chap10:cs24}
 {\Gamma_L \over \Gamma} \equiv
{\Gamma(B \rightarrow J/\psi K^{*})_L \over 
\Gamma(B \rightarrow J/\psi K^{*})}
 = {H_L \over H_L + H_T},
\end{equation}
\begin{equation} \label{chap10:cs25}
R_{\psi} \equiv 
{\Gamma(B \rightarrow J/\psi K^{*})\over 
\Gamma(B \rightarrow J/\psi K)} =
1.08 {(H_L + H_T) \over z^2 \left| 1 + 2 {C_1 \over a_2} \chi_{F_1} \right|^2}
\end{equation}
And the parity-odd (P-wave) transverse polarization measured 
in the transversity basis \cite{CLEO3,DDLR} is given:
\begin{eqnarray}\label{chap10:cs26}
|P_{\perp}|^2 &=& 
{|A_{\perp}|^2 \over |A_0|^2 + |A_{\parallel}|^2 +|A_{\perp}|^2}
= 2 c^2 y^2 
{ \left(1 + 2{c_1 \over a_2} \chi_{V} \right)^2 \over (H_L + H_T)}.
\end{eqnarray}

When $\chi_{F1} = \chi_{A1} = \chi_{A2} = \chi_{V} = \chi$,
a new factorization scheme comes appear. In this case, the nonfactorizable
terms only affect the coefficient $a_2$ as below:
\begin{eqnarray}\label{chap10:cs27}
a_2 \hspace{10mm} \longrightarrow \hspace{10mm}
 a_2^{eff} &=& a_2 \left( 1 + 2 {C_1 \over a_2} \chi \right) \nonumber \\
&=& a_2 + 2 C_1 \chi \nonumber \\
&=& \left( C_2 + {1 \over N_c} C_1 \right) + 2 C_1 \chi \nonumber \\
&=& C_2 + \xi C_1, \hspace{20mm} \xi = {1 \over N_c} + 2 \chi.
\end{eqnarray}
However the predictions of $\Gamma_L/\Gamma$, $R_{\psi}$, and $|P_{\perp}|^2$
in the mormal factorization method \cite{WBS} remain intact
since all nonfactorizable terms are cancelled out in
Eq.(\ref{chap10:cs24} - \ref{chap10:cs26}). 

\section{Phenomenological Model}

Let us see if the experimental measurements can be explained within 
the context of the factorization approach.
To proceed, we consider several phenomenological models of form factors:
\begin{enumerate}
\item 
The Bauer-Stech-Wirbel model (called BSW I here)\cite{WBS} 
in which $B \rightarrow K(K^{(*)})$ form factors 
are first evaluated at $q^2 = 0$ and
 then extrapolated to finite $q^2$ using a monopole type $q^2$-dependence
for all form factors $F_1,A_1,A_2$, and $V$ ;
\item
The modified BSW model (called BSW II here) \cite{NR}, 
takes the values of the form factors at $q^2 =0$ as in BSW I
but uses a monopole form factor for $A_1$ and a dipole form factor
for $F_1,A_2$,and $V$ ;
\item
The non-relativistic quark model by Isgur et al (ISGW)\cite{ISGW}
with exponential $q^2$ dependence for all form factors ;
\item
The model of Casalbuoni et al and Deandrea et al (CDDFGN)\cite{CDDFGN}
 in which the normalization at $q^2 =0$ is obtained in a model that
combines heavy quark symmetry with chiral symmetry for light vector degrees 
of freedom and also introduces light vector degrees of freedom.
Here all form factors are extrapolated with monopole behavior. 
\end{enumerate} 

Several authors have derived the $B \rightarrow K(K^{*})$ form factors 
from experimentally measured $D \rightarrow K(K^{*})$ form factors
at $q^2 = 0$ using the Isgur-Wise scaling laws based on the SU(2) heavy
quark symmetry \cite{HQS}, which are allowed to relate $B$ and $D$ form
factors at $q^2$ near $q^2_{max}$.
\begin{enumerate}
\item
The $ B \rightarrow K(K^{*})$ form factors are calculated 
in Ref.\cite{Keum1}
by assuming a constant for $A_1$ and $A_2$, 
a monopole type form factor for $F_1$, and dipole type for $V$.
\item
An ansatz proposed in Ref.\cite{Orsay}, which relies on ``soft'' Isgur-
Wise scaling laws and a monopole type for $A_1$ and a dipole type
for $A_2,V,F_1$.
\item
For Ref\cite{CT}, they are computed by advocating a monopole extrapolation
for $F_1,A_0, A_1$, a dipole behavior for $A_2, V$, and an approximately
constant for $F_0$. 
\end{enumerate}

Table 1 summerizes the predictions of $\Gamma_L/\Gamma, R_{\psi}$ and 
$|P_{\perp}|^2$ in above-mentioned various form factor models
within the factorization approach by assuming the absence of inelastic
final-state interactions. It appears that Keum's and CT's predictions 
are most close to the data.

Gourdin et al \cite{Gourdin2} also have suggested that the ratio
$R_{\eta_c} = {\cal B}(B \rightarrow \eta_c K^{*})/
{\cal B}(B \rightarrow \eta_c K)$ would provide a good test of the
factorization hypothesis in Class II decays. Using data of Particle Data
Group \cite{PDG} of ${\cal B}(B^{+} \rightarrow K^{+}J/\psi)
 = (1.02 \pm 0.14)\%$, we expect ${\cal B}(B^{+} \rightarrow K^{+}\eta_c)
= (1.14 \pm 0.31) \times 10^{-3}$, which could be within reach of near future
experimental data accumulation. Other ratios of decay rates in modes 
with charmonium mesons may also be used to test for the violation 
of factorization\cite{Keum1,Keum2}.

 
\begin{table}[b]
\caption{Experimental data and theoretical predictions
for $\Gamma_L/\Gamma, R_{\psi}$, and $ |P_{\perp}|^2$.}
\begin{center}
\begin{tabular}{|c||c|c|c|} \hline
 -  & $\Gamma_L/\Gamma$ & $R_{\psi}$ & $|P_{\perp}|^2$ \\   \hline \hline
 ARGUS \cite{ARGUS}  & $ 0.97\pm0.16\pm0.15$ & -  & -   \cr 
 CLEO II(95) \cite{CLEO2} & $0.80\pm0.08\pm0.05$  & $1.71\pm0.34$ & -  \cr
 CDF \cite{CDF1,CDF2}     & $0.65\pm0.10\pm0.04$  & 
$1.32\pm0.23\pm0.16$ & -  \cr
 CLEO II(96) \cite{CLEO3} & $0.52\pm0.07\pm0.04$ & $1.45\pm0.20\pm0.17$ &
$0.16\pm 0.08\pm0.04$ \\
\hline \hline
BSW I \cite{WBS} & 0.57 & 4.23 & 0.09 \cr
BSW II \cite{NR} & 0.36 & 1.61 & 0.24 \cr
ISGW \cite{ISGW} & 0.07 & 1.72 & 0.52 \cr
CDDFGN \cite{CDDFGN} & 0.36 & 1.50 & 0.30  \cr
JW \cite{JW}    & 0.44 & 2.44 &  \\
\hline
Orsay \cite{Orsay} & 0.45 & 2.15 & 0.25 \cr
Keum \cite{Keum1} &$ 0.59\pm0.07$ & $1.74\pm0.38$ & $0.14 \pm 0.05$ \cr
CT \cite{CT} & 0.56 & 1.84  & 0.16 \\
\hline
\end{tabular}
\end{center}
\end{table}

\end{document}